\documentclass[a4paper]{article}
\usepackage[bookmarks=false]{hyperref}
\usepackage{icrc2013}
\usepackage[english]{babel}
\title{Sites in Argentina for the Cherenkov Telescope Array Project}

\shorttitle{ICRC 2013 Template}

\authors{
Ingo Allekotte$^{1}$,
Gonzalo de la Vega $^{2}$,
Alberto Etchegoyen$^{3}$,
Beatriz Garc\'ia$^{2,4}$, 
Alexis Mancilla$^{2}$,
Javier Maya$^{2}$,
Diego Ravignani$^{3}$,
Adri\'an Rovero$^{5}$, 
for the CTA Consortium$^{6}$.
}

\afiliations{
$^1$ Centro At\'omico Bariloche and Instituto Balseiro (CNEA, UNCuyo), Bariloche, Argentina \\
$^2$ ITeDA-M, Mendoza, Argentina\\
$^3$ ITeDA (CNEA, CONICET, UNSAM), Buenos Aires, Argentina \\
$^4$ Universidad Tecnol\'ogica Nacional, Facultad Regional Mendoza, Mendoza, Argentina\\
$^5$ Instituto de Astronom\'ia y F\'isica del Espacio (CONICET-UBA), Buenos Aires, Argentina \\
$^6$ www.cta-observatory.org \\
\scriptsize{
}
}

\email{ingo@cab.cnea.gov.ar}

\abstract{The Cherenkov Telescope Array (CTA) Project will consist of two arrays of atmospheric Cherenkov
telescopes to study high-energy gamma radiation in the range of a few tens of GeV to beyond 100 TeV.
To achieve full-sky coverage, the construction of one array in each terrestrial hemisphere is considered.
Suitable candidate sites are being explored and characterized. The candidate sites in the Southern
Hemisphere include two locations in Argentina, one in San Antonio de los Cobres (Salta Province, Lat.
24:02:42 S, Long. 66:14:06 W, at 3600 m.a.s.l) and another one in El Leoncito (San Juan Province, Lat.
31:41:49 S, Long. 69:16:21 W, at 2600 m.a.s.l).

Here we describe the two sites and the instrumentation that has been deployed to characterize them. We
summarize the geographic, atmospheric and climatic data that have been collected for both of them.
}

\keywords{Gamma Ray Astronomy, Atmospheric Cherenkov Telescopes, site characterization.}

\begin{document}
\maketitle

\section{Introduction}

The objective of the Cherenkov Telescope Array Project (CTA) is to study high energy cosmic gamma radiation in the range from 20 GeV to above 100 TeV with high statistics and precision. It will consist of two observatories, one in each hemisphere, to achieve full-sky coverage. The Southern CTA Observatory will consist of approximately 100 telescopes of different sizes (6 m, 12 m, 24 m diameter mirrors) forming an array spread over an area of 10 km$^2$. 

Efforts are being undertaken to select the most suitable sites for the installation of the CTA Observatory. The sites have to fulfill a series of requirements of different kind: geographic (area, slope, altitude, etc.), atmospheric (low cloudiness and humidity, low light pollution, low aerosol content), accessibility and infrastructure (power provision, data connection , facilities, manpower, accessibility) and safety (earthquakes, hail, tornadoes, sand storms). Based on these requirements and on previous knowledge of potential sites in Argentina \cite{rovero}, two candidate sites were pre-selected by the Argentinian CTA Collaboration: 
\begin{itemize}
\item \textbf{"San Antonio de los Cobres (SAC)"} in Salta Province: Lat.
24:02:42 S, Long. 66:14:06 W, at 3600 m.a.s.l.
\item \textbf{"El Leoncito (LEO)"} in San Juan Province: Lat.
31:41:49 S, Long. 69:16:21 W, at 2600 m.a.s.l. 
\end{itemize}
  
In this article we describe these sites and their characteristics.

\section{San Antonio de los Cobres (SAC) Site}

\subsection{Site description}

The SAC site is located in the Argentinian Puna, i.e., the Eastern Andes Highland. It is 22 km North of the town of San Antonio de los Cobres (pop. 6000), 
capital of the Department of Los Andes. The town is 170 km from Salta City, a less than 3 hour drive along a mostly paved road. 

One of the main characteristics of the San Antonio site is its high altitude (3600 meters), 
advantageous for gamma ray detection at the lowest energy range of CTA \cite{melo}. 

The SAC candidate site is located on land owned by the Government of the Province of Salta, who has already expressed its willingness to provide it free of costs. 

 \begin{figure}[t]
  \centering
  \includegraphics[width=0.4\textwidth]{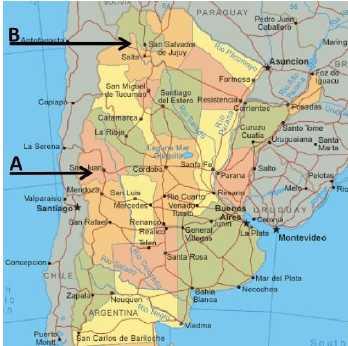}
  \caption{Location of the LEO and SAC sites, labelled "A" and "B" respectively.}
  \label{map-argentina}
 \end{figure}

\begin{figure}[t]
  \centering
  \includegraphics[width=0.4\textwidth]{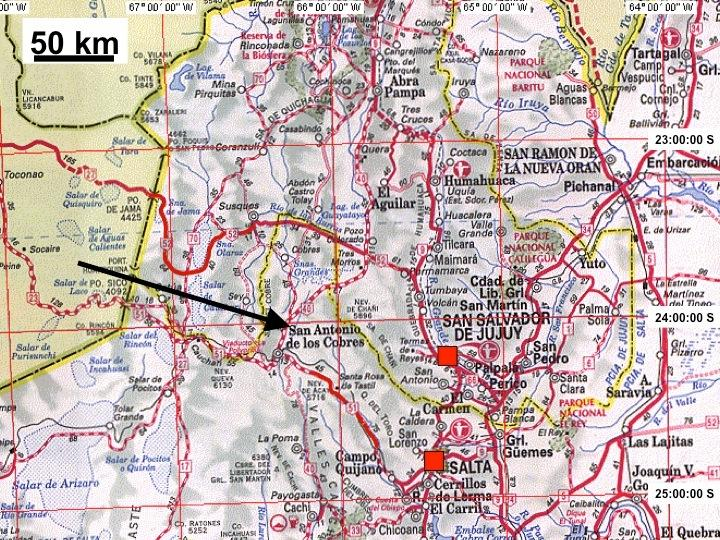}
  \caption{Road map around the SAC site, showing access roads and nearby towns and cities.}
  \label{map-sac}
 \end{figure}
 
\subsection{Geographical Characteristics}

The site consists of a flat circular area in excess of 10 km$^2$, with a gentle NW slope of less than 2$^\circ$. A detailed topographical survey has been done to assess the surface profile of the whole site and its surroundings. Soil studies \cite{viramonte} indicate that the ground is formed by deposits from alluvial fans, composed by non plastic materials, with predominance of poorly graded gravel or sandy gravel ("GP", according to the Unified Soil Classification System). Its value as foundation when not subject to frost action is considered good to excellent, the value as base directly under wearing surface is poor to fair; potential frost action none to very slight; compessibility and expansion almost none; dranaige characteristic excellent; Dry Unit Weight (pef): 120 -130; Field California Bearing Ratio of 35 to 60; subgrade modulus k (pei): 300 or more; MR range (ksi): 25-35; MR Default (ksi): 30. Georadar profilings to a depth of 12 to 15 m demonstrate the existence of a thick and uniform layer (7-8 m) of this kind of materials, which lie on what is considered the tertiary basement of the San Antonio riverbed. These preliminary studies conclude that the general geotechnical characteristics of the site are good to very good for the foundations of the CTA telescopes. 

The seismic activity in the Andes range is related to the subduction of the Nazca plate under the South American plate. For this reason, the depth of seisms increases from West to East, typical earthquakes occurring in the SAC area are at depths of 150 to 300 km. A site-specific seismic
assesment was done for the SAC site \cite{viramonte-sismos} in which it is concluded that peak ground accelerations with 10\% probability of exceedance in 50 years (mean return period of 475 years) is below 1.4 m/s$^2$ for horizontal acceleration, and at most 60\% thereof for vertical motion. 

A flood risk assessment for the proposed CTA sites in Argentina has also been done \cite{hydrogis} concluding that the flood risk from rivers at both sites is low.

\subsection{Climate and Atmosphere}

Historic meteorological data from the National Meteorological Service is available from a weather station operating in the town of San Antonio for more than 10 years. This Puna region is characterized as a semidesertic area, with low cloud coverage and few precipitations (below 100 mm per year), mainly concentrated in the December to March summer season. Additionally, an automated meteorological station acquired data in San Antonio from 2008 to 2011. 

A new, dedicated automated meteorological station (Davis Weather Monitor II) has been installed at the site in April 2011 and 
is recording data for the CTA candidate site (temperature, pressure, humidity, wind speed and direction). A 10 m pole was set up at the site to install the anemometer and for an antenna for remote data transfer. An additional communications antenna was set up at a nearby hill, with direct line of sight to the site antenna and to the Municipality of San Antonio, from where the data is transferred via internet to a server in Mendoza and made available online \cite{weather-sac}. 

Results from weather measurements indicate that SAC can occassionally get as cold as -10$^\circ$C during winter time, with maximum temperatures of 24$^\circ$C in summer. For the period of time recorded, windspeed is below the operations limit required for CTA (36 km/h) during over 99\% of the night time. Windspeed has never been recorded beyond the maximum allowed damage limit (120 km/h in a 10-minute average).

Also, to measure the sky background brigtness, a Sky Quality Meter (Unihedron SQM-LR with narrow field of view of 10$^\circ$ FWHM, with an SQM normal filter centered around the Johnson V band) was installed at the site in April 2011. This device was previously cross-calibrated in the laboratory with other similar instruments. Data is being provided together with the data from the weather station.  To interpret the obtained data, the contribution of stars to the night sky background has to be taken into account, as the Galactic plane is well within the field of view of the instrument for a considerable fraction of the time. Results indicate an average sky background brightness of 21.9 mag/arcsec$^2$, well within the CTA Project requirements. 

For cloudiness and background light studies, a full-sky camera was designed and developed for CTA and installed at the SAC site in December 2011 \cite{ebr}. The system has been accummulating data for several months (with some interruptions) and this data is now being processed. Preliminary results indicate low cloud coverage during Austral spring 2012, as expected. 

A weather and atmospheric monitoring instrument, called ATMOSOCOPE, was specifically designed for CTA. One of these is deployed at the SAC site since August 2012 \cite{bulik}.
Weather data from the Atmoscope, the Davis meteorological station and the SQM have been cross-checked and found to be in good agreement. 

Due to the scarcity of precipitations, snowfall is not a problem for the SAC site. Although occassional ice precipitation, particularly snow pellets, has been reported as very rare events in other areas of the Puna Highland, there are no systematic records available. Thunderstorms occur mostly during the summer season, so lightning protection would be required for the CTA equipment.

To determine the aerosol content of the atmosphere, a test measurement with an aerosol spectrometer \cite{piacentini} (Grimm Model 1.109) has been undertaken, and particulates collected during one hour were analyzed both in quantity and aerosol size. The low aerosol levels thus determined are consistent 
with the results provided by the MODIS satellite \cite{modis} for Aerosol Optical Depth at 550 nm green band wavelength. Modellings of the site conditions foresee no problems with dust storms, as there are no large dust sources nearby the SAC site.

Recently, test mirrors have been installed on the site to study water condensation and resistance to the environment \cite{medina}. 

\subsection{Infrastructure}

As capital of the Department, the town hosts some infrastructure to provide services to the town itself and to different mining companies that operate in the Western Highland of the Salta Province. A local hospital provides medical assistance for emergencies and lower risk interventions. One higher rank hotel and a few restaurants provide services in town. A high-voltage power line which connects Salta to Chile passes nearby the town; the installation of a transformer and a medium-voltage extension is foreseen by the Province of Salta. The whole Province, including San Antonio, is being interconnected with optical fiber and connection to the Argentinian internet grid is foreseen for the near future. A railway which connects Salta City to Antofagasta in Chile and allows a stop in San Antonio provides a mean of transportation for heavy and bulky loads. Also, a wide and modern route connecting Jujuy to Antofagasta through the Jama Pass passes only 50 km North of the site and can be used for lorries and trucks. 

The recently approved and funded LLAMA Project \cite{llama} is constructing a large radio-antenna, which will measure interferometrically with ALMA in the Atacama desert. LLAMA will be deployed in Alto Chorrillos, a site at 28 km from San Antonio de los Cobres town and only 30 km to the SAC site. Synergy between both LLAMA and the CTA Project can be expected in case the SAC site is chosen for CTA.

\section{El Leoncito (LEO) Site}

\subsection{Site description}

The site is located nearby the CASLEO Observatory (www.casleo.gov.ar), on a plateau 
extending more than 15 km in the North-South direction and 
3 to 5 km in the East-West direction. 
Although it presents a slight East to West slope of nearly 4$^\circ$ and some 
dry channels, its topography is acceptable for the CTA requirements. The site 
is located in the El Leoncito National Park, which was created primarily to 
preserve the clear skies and atmosphere around CASLEO Observatory.

\begin{figure}[t]
  \centering
  \includegraphics[width=0.4\textwidth]{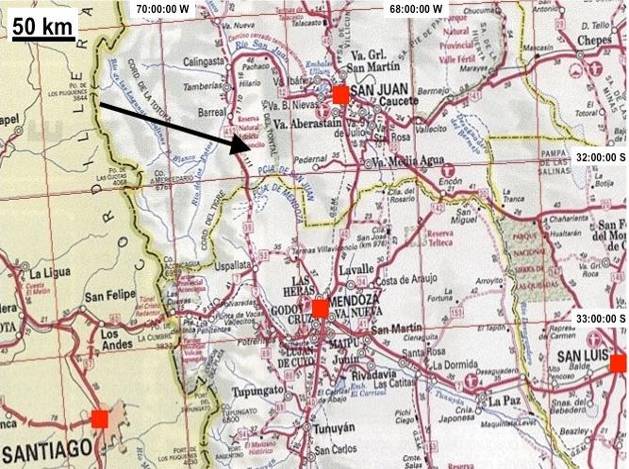}
  \caption{Road map around the LEO site, showing access roads and nearby towns and cities.}
  \label{leo-map}
 \end{figure}
 
\subsection{Geographical Characteristics}

A geophysical profiling was performed for the site \cite{ruiz} using 
gravimetric, magnetic, geodesic and seismic measurement methods, which 
allowed to understand the vertical profile of the soil. The upper 
layer is mostly composed of alluvial sedimentary material.

To better characterize the seismic activity in the region, a recent dedicated study \cite{triep} has been performed. It concludes that the maximum peak ground acceleration at the site, with a 10\% probability of exceedance in 50 years, is 2.7 m/s$^2$, therefore well within the CTA requirements. 

\subsection{Climate and Atmosphere}

With the CASLEO Observatory running for over 20 years, a huge amount of climate and atmospheric data is available from ground measurements performed over many years. 
Temperature range is from -11$^\circ$C to 32$^\circ$C, and windspeeds exceeding the CTA operational limit have ben recorded for less than 1\% of the night time.

Cloud coverage has been recorded for many years \cite{martinis} using a ground based all-sky imaging (ASI) camera provided by Boston University (www.buimaging.com) installed at CASLEO Observatory. According to this study, and with data collected between 2006 and 2010, 78\% of the night time the skies are completely clear of clouds. More recently, a dedicated full sky camera has also been installed by the CTA Collaboration at the proposed site for CTA \cite{ebr}, data is currently being analyzed.  

The CASLEO Observatory has also systematically measured the night sky background brightness with an SQM device. An additional SQM instrument has been deployed at the center of the candidate CTA location. Results show that the site is within CTA specifications once the luminosity from the Galactic Center and the Milky Way are subtracted when in the field of view of the instrument. 

An ATMOSCOPE, similar to the one installed at SAC, has also been installed in January 2012 and records temperature, humidity, wind and night sky brightness. Results are consistent with those provided by the automated weather stations.  

Aerosols have been studied (both in quantity and size distribution) with an AERONET  (AErosol RObotic NETwork)  which is a network of ground-based remote sensing aerosol monitors established by NASA and PHOTONS  to measure atmospheric transmittance \cite{aeronet}. Level 2 
quality data from years 2011-2012 show that the cleanliness of the atmosphere is 
comparable to the best sites in the world \cite{ristori}.

Measurements with the Grimm spectrometer have also been performed at LEO, data collected during two days (Dec. 27 and 28, 2012) show distributions and densities similar to those measured at the Auger Observatory	in the Andes range (i.e., typical distributions for semi-­arid Andes regions).	
  
\subsection{Infrastructure}

The nearby CASLEO Observatory is an astronomical 
facility inaugurated in 1986, whose main instrument is a 2.15 m telescope. 
Therefore, the basic facilities are already available (electrical power from the national grid, 
office space, workshops, lodging and restaurant). Although the power provision is not sufficient for the CTA requirements, an extension of the power line from the nearby town of Calingasta is already foreseen. This line can also carry the optical fiber required for communications. 

The LEO site is at a distance of 30 km to El Barreal, a small touristic village (pop. 4,000) 
with availability of all basic services (hospital, ATM, police, tourism office, supermarkets, 
hotels, restaurants, gasoline station, stores), which makes it an ideal town of residence for staff personnel. 
The site is at 250 km (less than 3 hours) from San Juan City (pop. 450,000), 
and nearly the same distance from Mendoza City (pop 800,000). 
Both are highly developed province capital cities.

\section{Conclusions}

Both the El Leoncito and San Antonio de los Cobres sites are suitable candidate sites for the installation of the CTA Observatory. Efforts are ongoing to better characterize these sites, to provide a complete understanding of their advantages and mitigate their disadvantages. 

Regional and national support both at scientific and political level are provided for both sites. With the successful commissioning and operation of the Pierre Auger Observatory in Argentina (www.auger.org), valuable experience has already been gained in handling costs, importations, hiring of personnel and construction of big facilities.

\vspace*{0.5cm}
\footnotesize{{\bf Acknowledgments:}{We gratefully acknowledge support from the following agencies and organizations:
Ministerio de Ciencia, Tecnolog\'ia e Innovaci\'on Productiva (MinCyT),
Comisi\'on Nacional de Energ\'ia At\'omica (CNEA) and Consejo Nacional  de
Investigaciones Cient\'ificas y T\'ecnicas (CONICET) Argentina; State Committee
of Science of Armenia; Ministry for Research, CNRS-INSU and CNRS-IN2P3,
Irfu-CEA, ANR, France; Max Planck Society, BMBF, DESY, Helmholtz Association,
Germany; MIUR, Italy; Netherlands Research School for Astronomy (NOVA),
Netherlands Organization for Scientific Research (NWO); Ministry of Science and
Higher Education and the National Centre for Research and Development, Poland;
MICINN support through the National R+D+I, CDTI funding plans and the CPAN and
MultiDark Consolider-Ingenio 2010 programme, Spain; Swedish Research Council,
Royal Swedish Academy of Sciences financed, Sweden; Swiss National Science
Foundation (SNSF), Switzerland; Leverhulme Trust, Royal Society, Science and
Technologies Facilities Council, Durham University, UK; National Science
Foundation, Department of Energy, Argonne National Laboratory, University of
California, University of Chicago, Iowa State University, Institute for Nuclear
and Particle Astrophysics (INPAC-MRPI program), Washington University McDonnell
Center for the Space Sciences, USA. The research leading to these results has
received funding from the European Union's Seventh Framework Programme
([FP7/2007-2013] [FP7/2007-2011]) under grant agreement Nr. 262053.}}

\end{document}